\documentclass[a4paper,12pt]{article}
\usepackage{setspace}
\newcommand{\bpsi}{\bar{\psi}}
\newcommand{\bPsi}{\bar{\Psi}}
\newcommand{\bsigma}{\bar{\sigma}}
\newcommand{\dpsi}{\psi^{\dagger}}

\newcommand{\Hc}{{\cal H}} \hoffset-1in \textwidth15.6cm \textheight24.3cm
\begin{document}

\begin{center}
{\Large\bf Generation of Neutrino Mass in a Kalb-Ramond Background in Large Extra Dimensions }\\[20mm]
Soumitra SenGupta\footnote{E-mail: tpssg@iacs.res.in}\\
{\em Department of Theoretical Physics\\
 Indian Association for the Cultivation of Science\\
Calcutta - 700032, India}\\ \vskip .2cm
Aninda Sinha \footnote{E-mail:A.Sinha@damtp.cam.ac.uk } \\
{\em Department of Applied Mathematics and Theoretical Physics,\\
Wilberforce Road, Cambridge CB3 0WA, UK}\\[20mm]
\end{center}

\vspace{0.5cm} {\em PACS Nos.: 04.20.Cv, 11.30.Er, 12.10.Gq}
\vspace{0.5cm}

\begin{abstract}
In this paper we investigate whether spacetime torsion induced by
a Kalb-Ramond field in a string inspired background can generate
a mass for the left-handed neutrino. We consider an
Einstein-Dirac-Kalb-Ramond lagrangian in higher dimensional
spacetime with torsion generated by the Kalb-Ramond antisymmetric
field in the presence of a bulk fermion. We show that such a
coupling can generate a mass term for the four dimensional
neutrino after a suitable large radius compactification of the
extra dimensions.
\end{abstract}
\onehalfspacing
\section{Introduction}
Ever since Einstein-Cartan theory was proposed, space-time torsion
has been considered as an integral part of any gravitational
theory where the background geometry is not only characterized by
curvature but also by an asymmetric part of the affine connection
called space-time torsion\cite{hehl}. Just as mass-energy is known
to be the source of curvature in spacetime, torsion originates
from spin in space-time\cite{saba}. In spite of its generality
over Einstein's theory, a theory with torsion didn't draw that
much attention because of the lack of experimental support
possibly due to a weak value of the torsion. Moreover, it was
shown that the torsion field, even if it exists, cannot couple to
the electromagnetic field in a gauge invariant way. Thus there is
no possibility of finding any signal of torsion in electromagnetic
experiments.

There was a revival of interest in the subject after the advent of
string theory\cite{gsw} where the field strength corresponding to
the massless second rank antisymmetric tensor field (known as
Kalb-Ramond field) in the heterotic string spectra was identified
as the background spacetime torsion. The remarkable feature in
such a theory was that now one could couple electromagnetisnm with
torsion preserving the $U(1)$ gauge invariance\cite{pmssg}. The
Chern-Simons term needed to cancel gauge anomaly compensates the
gauge symmetry violating term originating from the torsion
coupling. Such a theory offers possible explanations for various
phenomena like the optical activity purportedly observed in the
electromagnetic radiation coming from distant galactic
sources,\cite{skpmssgasss} presently observed accelerating phase
of the Universe and others\cite{ssgss}. As torsion plays a crucial
role in such explanations, one is tempted to propose these
observations as indirect support for string theory as well as for
the existence of spacetime torsion.

It has further been shown in a recent work that while torsion and
curvature effects both have the same coupling with other matter
fields at the Planck scale in higher dimensions, the torsion
coupling becomes weak as the extra dimensions are
compactified\cite{rs} via the Randall-Sundrum
mechanism\cite{bmssssg}. All these motivated us to explore
whether torsion in extra dimensions could provide an explanation
for the small neutrino mass predicted from neutrino oscillation
experiments. We have shown earlier that the torsion coupling may
indeed result in the helicity flip\cite{ssgas} in massive
neutrinos which may suggest a possible explanation for the solar
neutrino anomaly. The origin of such a mass term has been explored
in several works from various points of
view\cite{Dienes,Mohapatra}. In particular, it has been argued
that bulk fermions living in higher dimensions could couple to
fermionic fields living in 3+1 dimensions, for instance via the
Higgs field, and generate mass\cite{addmass,kddug}. Here we take a
new path with the input that torsion resides in the bulk and
interacts with the standard model fermion on the wall of the brane
as all the standard model particles are assumed to reside on the
brane. Such an approach has special significance in the context
of string theory where the second rank antisymmetric Kalb-Ramond
field may act as the source of torsion in the background. In fact
the KR field strength and torsion can be equated as we shall
explain shortly. We thus explore the possibility of generating
neutrino mass from the geometric property of the background
spacetime. In such a scenario we compactify the extra dimensions
following the scheme proposed by Arkani Hamed, Dimopolous and
Dvali(ADD)\cite{add}. Such a compactification scheme involves
large extra dimensions which pull down the Planck scale near the
electroweak scale thereby solving the so called hierarchy
problem. In such a scenario all the standard model fields are
confined on the brane where gravity resides in the bulk. The
torsion field being an integral part of the geometry of spacetime
like gravity is also assumed to reside in the bulk. The
compactified field strength of the Kalb-Ramond field induced
torsion in 4D is known to be related with the string axion by the
well known duality relation. This axion is assumed to have frozen
into its vacuum expectation value during a much earlier epoch and
is responsible for inducing a mass term for the left handed
neutrino on the brane as well as contributions to the massive
towers of Kaluza-Klein modes of the bulk fermion.

We briefly recall the salient features of the ADD type models. In
such models \cite{add} , the compact and Lorentz degrees of
freedom can be factorized. The string scale $M_s$ (which can be
as low as tens of TeV) controls the strength of gravity in $(4+n)$
dimensions, and is related to the 4-dimensional Planck scale
$M_P$ by
\begin{equation}
\frac{R^n}{M_P^2} = (4\pi)^{n/2}\Gamma(n/2) M_s^{-(n+2)}
\end{equation}

\noindent where  R is the compactification radius. The current
limits on the departure from Newton's law of gravity at small
distances are consistent with $R$ within a $mm$, for $n \ge 2$.
On compactifying the extra dimensions we get a tower of
Kaluza-Klein (KK) modes on the brane where we reside. Thus a
massless field in the bulk in general gives rise to a massive
spectrum and  the density of states is given by
\begin{equation}
\rho({m}_{\vec{n}})=\frac{R^n
{{m}_{\vec{n}}^{(n-2)}}}{(4\pi)^{n/2}\Gamma(n/2)}
\end{equation}
where ${{m}_{\vec{n}}}={(\frac{4\pi^2{\vec{n}}^2}{R^2})}^{1/2}$
is the mass of a KK state with $\vec{n}=(n_1,n_2,........,n_n)$
\cite{han}. Consequently, in any process (involving the graviton,
for example) where a cumulative contribution from the tower is
possible, a summation over the tower of fields, convoluted by the
density, causes an enhancement, in spite of the suppression of
individual couplings by $M_P$. One thus expects appreciable
contributions to various processes at energies close to $M_s$.

In the scenario adopted by us, the source of torsion is taken to
be the rank-2 antisymmetric Kalb-Ramond (KR) field $B_{MN}$ which
arises as a massless mode in heterotic string theories\cite{gsw}.
To understand the above statement, let us recall that the low
energy effective action for the gravity and  Electromagnetic
sectors in D dimensions is given by

\begin{equation}
S = \int~ d^{D}x \sqrt{-G} ~\left[~R(G) ~-~
    \frac{1}{4} F_{MN}F^{MN} ~+~
    \frac{3}{2} H_{MNL}H^{MNL} ~\right]
\end{equation}

It has been shown earlier \cite{pmssg} that an action of the form

\begin{equation}
S = \int~ d^{D}x \sqrt{-G} ~\left[~R (G,T) ~-~
    \frac{1}{4} F_{MN} F^{MN} ~-~
    \frac{1}{2}H_{MNL} H^{MNL} ~+~
     T_{MNL}H^{MNL}\right]
\end{equation}

\noindent reproduces the low energy string effective action if
one eliminates the torsion field $ T_{MNL}$ ( which is an
auxiliary field) by using the equation of motion $ T_{MNL} ~=~
H_{MNL}$.

Thus torsion can be identified with the rank-3 antisymmetric
field strength tensor $H_{MNL}$ which in turn is related to the
KR field $B_{MN}$\cite{kr} as

\begin{equation}
H_{MNL} = \partial_{[M}B_{NL]}
\end{equation}

\noindent with each Latin index running from $0$ to $4$ in a
five-dimensional theory. (Greek indices, on the other hand, run
from 0 to 3.) Furthermore, we use the KR gauge fixing conditions
to set $B_{4\mu}~=~0$. Therefore, the only non-vanishing KR field
components correspond to the brane indices. These components, of
course, are functions of both compact and non-compact
co-ordinates.

For a spin-1/2 fermion in a spacetime with torsion, the extended
Dirac Lagrangian density is given by \cite{aud,fig}:
\begin{equation} {\cal L}_{fermion} ~=~ \bar{\psi}~\left[i
\gamma^{\mu}~\left(
\partial_{\mu} ~-~ \sigma^{\rho \beta} v^{\nu}_{\rho} g_{\lambda
\nu} \partial_{\mu} v^{\lambda}_{\beta} ~-~ g_{\alpha \delta}
\sigma^{a b} v^{\beta}_a v^{\delta}_b \tilde{\Gamma}^{\alpha}\,
_{\mu \beta} \right) \right]~\psi \end{equation} where
~$v^{\mu}_a$~ denote the tetrad connecting the curved space with
the corresponding tangent space and
\begin{equation}
\tilde{\Gamma}^{\mu}\,_{\nu\rho}=\Gamma^{\mu}\,_{\nu\rho}+H^{\mu}\,_{\nu\rho},
\end{equation}
$\Gamma$ denoting the connection without torsion.

The paper is organized as follows. In section 2, we consider a
five-dimensional theory with torsion and compactify on a circle.
Using the well known four dimensional duality relating the
torsion field to the axion, we show that the axion vev induces a
mass term for the neutrino. We generalize this mass term in
higher dimensions and show that for six large extra dimensions,
the value for the mass is of the order of a few electron-volts.
We conclude in section 3.

\section{Neutrino mass from torsion}
As in \cite{kddug}, let us start with a five-dimensional theory
and consider a compactification of the fifth dimension on a
circle of radius $R$. A five-dimensional massless fermion $\Psi$
can be decomposed into $(\psi_1,\psi_2)$ where $\psi_1,\psi_2$
are two component spinors. We consider the left-handed neutrino
$\nu_L$ to be moving in four dimensions only. Consider the
following effective action representing the kinetic terms for
$\Psi$ and $\psi$,
\begin{equation}
\int d^4x dy
\left[i\bPsi\gamma^{\mu}D_{\mu}\Psi+i\bPsi\gamma^{5}D_{5}\Psi
\right]+\int d^4x
dy\left[i\bpsi\gamma^{\mu}D_{\mu}\psi\right]\delta(y),
\end{equation}
where the four-dimensional $\psi=(\nu_L,N|_{y=0})$, $N=1/\sqrt{2}
(\psi_1+\psi_2)$ containing the right-handed neutrino for reasons
to be made clear later. We use the chiral representation where the
gamma matrices are given by,
\begin{equation}
\gamma^0=\pmatrix{0 & {-\mathbf 1}\cr {-\mathbf 1}& 0}\quad
\gamma^i=\pmatrix{0 & \sigma^i\cr -\sigma^i & 0}
\end{equation}
and $\gamma^5={\rm diag}({\mathbf 1,\mathbf -1})$. Using these,
\begin{eqnarray}
\int d^4 x dy\,
\{\dpsi_{1}i\sigma^{\mu}D_{\mu}\psi_1+\dpsi_{2}i\bsigma^{\mu}D_{\mu}\psi_2
-\dpsi_{2}iD_{5}\psi_1+i\dpsi_1 D_5 \psi_2\}+\\ \nonumber\int d^4x
dy \delta(y) \{\nu^{\dagger}_Li\sigma^{\mu}D_\mu
\nu_L+iN^{\dagger}\bsigma^{\mu}D_{\mu}N\}.
\end{eqnarray}
Now let us introduce space-time torsion generated by the
antisymmetric 3-form as explained above. We will get from the
connection piece terms like
\begin{equation}
\int d^4x dy{1\over M_s^{3/2}} \{i\bPsi
\gamma^{\mu\sigma\lambda}H_{\mu\sigma\lambda}\Psi +i\bpsi
\gamma^{\mu\sigma\lambda}H_{\mu\sigma\lambda}\psi\delta(y)\}
\end{equation}
where we have made the gauge-choice $B_{4\mu}=0$.

Using the four-dimensional duality relation for the massless
antisymmetric tensor mode,
\begin{equation}
H_{\mu\nu\rho}=\epsilon_{\mu\nu\rho\lambda}\partial^{\lambda} \chi
\end{equation}
where $\chi$ is the axion-field and is used in explaining the
strong CP problem via the PQ mechanism, we get using the standard
gamma-matrix relation $\epsilon_{\mu\nu\rho\lambda}
\gamma^{\mu\nu}=-2 i\gamma^{5}\gamma^{\rho\lambda}$ {\footnote{Our
conventions are such that
$\gamma^{\mu_1\mu_2\cdots}=\gamma^{\mu_1}\gamma^{\mu_2}\cdots,$
when $\mu_1\neq\mu_2\cdots$ and vanishes otherwise.}}
\begin{equation}
{1\over M_s^{3/2}}
\{\bPsi\gamma^{5}\gamma^{\rho}\partial_{\rho}\chi
\Psi+\bpsi\gamma^{5}\gamma^{\rho}\partial_{\rho}\chi \psi\}.
\label{massterm}
\end{equation}
Further we expand $\Psi=(\psi_1, \psi_2)$ in Kaluza-Klein modes,

\begin{eqnarray}
\psi_1(x,y)&=&\frac{1}{\sqrt{2\pi R}}\sum_{n}\psi_1(x)e^{iny/R} \\
\psi_2(x,y)&=&\frac{1}{\sqrt{2\pi R}}\sum_{n}\psi_2(x)e^{-iny/R}.
\end{eqnarray}

Also let us define the following linear combinations
$N(x,y)=1/\sqrt{2}(\psi_1+\psi_2),M(x,y)=1/\sqrt{2}(\psi_1-\psi_2)$
whose Kaluza-Klein modes are defined as
\begin{eqnarray}
N^{(n)}&=&\frac{\psi_1^{(n)}+\psi_2^{(n)}}{\sqrt{2}} \\
M^{(n)}&=&\frac{\psi_1^{(n)}-\psi_2^{(n)}}{\sqrt{2}}.
\end{eqnarray}

Using these in equation (\ref{massterm}) and integrating by parts,
results in mass terms
\begin{equation}
\sum_{n=1}^{\infty}n\,m N^{\dagger(n)}N^{(n)}+n\,m
M^{\dagger(n)}M^{(n)}+n\,m\nu_L N^{(n)}
\end{equation}
where,
\begin{equation}
m=\frac{\langle\chi\rangle R^{1/2}}{(2\pi M_s R)^{3/2}}.
\end{equation}

we have the following four-dimensional lagrangian,

\begin{equation}
{\cal L}={\cal L}_{KE}
+\sum_{n=1}^{\infty}(mn+\frac{n}{R})N^{(n)\dagger}N^{(n)}+(mn-\frac{n}{R})
M^{\dagger(n)}M^{(n)}\}
+\{\sum_{n=1}^{\infty}m_N^{(n)}\nu_L^{\dagger} N^{(n)} +{\rm
h.c.}\}
\end{equation}
where
\begin{equation}
m_{N}^{(n)}=n m
\end{equation}
Thus, there is no mass term for the zero-mode for $N$. We note
here that in 4 dimensions, $\langle \chi \rangle \approx f_{PQ}$
where $f_{PQ}$ is the Peccei-Quinn scale. When the axion lives in
$4+n$ dimensions, $f_{PQ}\sim \langle\chi\rangle R^{n/2}$ so that
for our five-dimensional model, the numerator in our mass-formula
is nothing but $f_{PQ}$ . If an axion is a boundary field confined
in 4 dimensions, the PQ scale $f_{PQ}$ is bounded by $M_s$. To
obtain a higher PQ scale which is demanded on astrophysical
grounds, the axion field has to be a bulk field\cite{Chang}. If
one considers a more general model with $n$ extra dimensions with
the axion field moving in a $p$ dimensional submanifold, the above
formula generalizes to
\begin{equation}
m=\frac{f_{PQ}}{(2\pi M_s R)^{(n+2)/2}}.
\end{equation}
where $f_{PQ}$ depends on $p$. Thus in the basis,
\begin{equation}
(\nu_L,N^{(1)},N^{(2)}\cdots),
\end{equation}
we get the following mass-matrix,
\begin{equation}
{\cal M}=\pmatrix {0 & m & 2m &3m  &\cdots \cr m & m+1/R&
0&0&\cdots \cr 2m &0&2m+2/R&0&\cdots\cr \vdots},
\end{equation}
The eigen-values of the above matrix can be computed numerically.
However one can make the following approximation. We consider the
case $m \ll 1/R$, and note that the total number of modes is
$O(M_s R)$. This results in the following $M_s R\times M_s R$
mass-matrix,
\begin{equation}
{\cal M}=\pmatrix {0 & m & 2m &3m  &\cdots \cr m & 1/R& 0&0&\cdots
\cr 2m &0&2/R&0&\cdots\cr \vdots}, \label{massmatrix}
\end{equation}
The eigen-values of the above matrix contain the masses of the
left-handed neutrino and other Kaluza-Klein modes. Since such a
matrix leads to a see-saw mechanism between $\nu_L$ and various
components of $N$, it is natural to assume that $N$ contains the
right-handed neutrino. The characteristic equation whose roots
yield the eigen-values for the mass-matrix in equation
(\ref{massmatrix}) is
\begin{equation}
\prod_{k=1}^{M_s R}(\lambda-{k\over
R})(\lambda-m^2\sum_{k=1}^{k=M_s R}\frac{k^2}{\lambda-{k\over
R}})=0.
\end{equation}
The above equation can be re-written as
\begin{eqnarray}
\lambda-m^2[-{1\over 2}M_s R^2(1+2 \lambda R +M_s
R)+\lambda^2R^3({\rm PolyGamma}(0,1-\lambda R)\\ \nonumber-{\rm
PolyGamma}(0.1-\lambda R+M_s R))]=0 \label{ana}
\end{eqnarray}
where PolyGamma$(0,x)=\Gamma'[x]/\Gamma[x]$. Equation (\ref{ana})
is a transcendental equation which can be solved graphically.
Instead of adopting a brute-force approach, we note that the
eigen-values of the above matrix can be computed by following the
procedure. Noting that usually $m\ll 1/R$, we decompose the
mass-matrix into
\begin{equation}
{\cal M}=\pmatrix {0 & &\cdots \cr 0 & 1/R&\cdots \cr 0
&0&2/R&\cdots\cr \vdots}+\pmatrix{0&m&2m&\cdots\cr m&0&\cdots\cr
2m&0&0&\cdots\cr\vdots}=\Hc_0+\Hc'.
\end{equation}
Now we treat $\Hc'$ as a perturbation and use standard
perturbative techniques which yield corrections to the
eigen-values of $\Hc_0$. To second order we have,
\begin{equation}
E=E_0+<0|\Hc'|0>+\sum_i\frac{|<i|\Hc'|0>|^2}{E_0-E_i}.
\end{equation}
This yields for the neutrino mass, \begin{equation}
m_{\nu_L}=m^2R(RM_s)^2,
\end{equation}
which comes as a second order correction, the first order
correction vanishing. Let us note here that a similar enhancement
to $m$ from the Kaluza-Klein modes was also obtained in
\cite{kddug} but there the enhancement had a logarithmic
dependence on $RM_s$ instead of the quadratic dependence that we
have here. As such the Kaluza-Klein enhancement in our model is
substantially more than the one in \cite{kddug}.

The values of $f_{PQ}$ have been computed for a variety of cases
in \cite{Chang} with $n$ extra-dimensions and $p$ being the number
of dimensions in which the axion field moves. Using these results,
one gets the following values for neutrino masses. For
$M_s=1~TeV$: \vskip .5cm
\begin{tabular}{|p{2cm}|p{3cm}|p{3cm}|p{3cm}|}
\hline $(p,n)$ & $R^{-1}(GeV)$ & $f_{PQ} (GeV)$ & $m_{\nu_L}(eV)$ \\
\hline
(1,2)& $10^{-13}$ &$10^{10}$& $10^{9}$ \\
\hline (2,4) & $10^{-5}$ &$10^{10}$ &$10^{2}$
\\ \hline
(3,6)& $10^{-3}$ & $10^{13}$&$10 $ \\
\hline
\end{tabular}
\vskip 1cm

Curiously, the case of most interest is the one with 6 extra
dimensions which is what one would expect when the starting point
is a superstring theory. If we are to believe that torsion does
explain the existence of massive neutrinos, a theory with just 2
extra dimensions would lead to a widely different value for the
neutrino mass {\footnote{The determination of the absolute value
of individual neutrino masses is an ongoing experimental problem
(See for example, \cite{Pas}.)}}. However we note here that for
this case, $m\approx 10^{-14}GeV$ and $1/R\approx 10^{-13}GeV$ as
a result of which the approximation that we have used {\it may}
break down. For the other two cases, the condition $m\ll 1/R$ is
indeed satisfied. We also note here that the $(n,p)=(2,1)$ case
seems to be ruled out on astrophysical grounds \cite{Chang}. In
fact, it seems that the number of dimensions $p$ in which the
axion lives in has to be such that $p\geq 2$. Thus the absurdly
large value for the case with two extra dimensions should not be
a cause for concern.
\section{Conclusion}
In this paper we have shown that large extra dimensions,
originally proposed to resolve the gauge hierarchy problem,
provides a possible mechanism to generate mass for neutrino when
compactified  following the ADD scheme. Spacetime torsion turns
out to play the crucial role for this purpose. The bulk fermion
communicates with the brane neutrino through the axion (dual to
the Kalb Ramond induced torsion)coupling and the resulting axion
vev determines the neutrino mass in 4D. We have determined the
dependence of this mass on the number of extra dimensions where
the axion resides. The magnitude of the mass turns out to match
well with the present bound on neutrino mass for a certain choice
of the number of extra large dimensions. This choice seems to be
consistent if the underlying fundamental theory was a superstring
theory. Thus apart from gravity, another geometric feature of
spacetime namely the torsion, in the bulk may generate mass for
neutrino at a scale where the axion freezes into it's vev
following a large scale compactification.

\section*{Acknowledgements}{\small { SSG acknowledge support from Board of Research in Nuclear sciences,
Government of India, under grant no. 98/37/16/BRNS cell/676. AS is
supported by the Gates Cambridge scholarship and the Perse
scholarship of Gonville and Caius College, Cambridge. }}

\end{document}